%% file: aaai25.tex
\documentclass[letterpaper]{article} 
\usepackage{aaai25}  
\usepackage{times}  
\usepackage{helvet}  
\usepackage{courier}  
\usepackage[hyphens]{url}  
\usepackage{graphicx} 
\usepackage{natbib}  
\usepackage{caption} 
\usepackage{algorithm}
\usepackage{algorithmic}
\usepackage{array}
\usepackage{multirow}
\usepackage{booktabs}
\usepackage{subfigure}
\usepackage{xcolor}

\usepackage{newfloat}
\usepackage{listings}
\DeclareCaptionStyle{ruled}{labelfont=normalfont,labelsep=colon,strut=off} 
\floatstyle{ruled}
\newfloat{listing}{tb}{lst}{}
\floatname{listing}{Listing}
\title{Enhancing Circuit Trainability with Selective Gate Activation Strategy}
\author{
    JeiHee Cho\textsuperscript{\rm 1},
    Junyong Lee\textsuperscript{\rm 1},
    Daniel Justice\textsuperscript{\rm 2},
    Shiho Kim\textsuperscript{\rm 1}
}
\affiliations{
    \textsuperscript{\rm 1}Yonsei University,~\textsuperscript{\rm 2}Carnegie Mellon University\\ jeiheec@gmail.com
}

\usepackage{bibentry}

\begin{document}

\maketitle

\begin{abstract}
Hybrid quantum-classical computing relies heavily on Variational Quantum Algorithms (VQAs) to tackle challenges in diverse fields like quantum chemistry and machine learning.  However, VQAs face a critical limitation: the balance between circuit trainability and expressibility.  Trainability, the ease of optimizing circuit parameters for problem-solving, is often hampered by the Barren Plateau, where gradients vanish and hinder optimization. On the other hand, increasing expressibility, the ability to represent a wide range of quantum states, often necessitates deeper circuits with more parameters, which in turn exacerbates trainability issues.
In this work, we investigate selective gate activation strategies as a potential solution to these challenges within the context of Variational Quantum Eigensolvers (VQEs). We evaluate three different approaches: activating gates randomly without considering their type or parameter magnitude, activating gates randomly but limited to a single gate type, and activating gates based on the magnitude of their parameter values. Experiment results reveal that the Magnitude-based strategy surpasses other methods, achieving improved convergence. 
\end{abstract}

%

\section{Introduction}
\input{intro}

\section{Main}
\input{main}

\section{Experiments}

\input{exp}

\section{Conclusion and Future Direction}
\input{conclusion}

\section{Acknowledgements}
This research was supported by Quantum Computing based on Quantum Advantage challenge research(RS-2023-00257561) through the National Research Foundation of Korea(NRF) funded by the Korean government (Ministry of Science and ICT(MSIT)).

\end{document}

%% file: intro.tex
Quantum computing has shown promise in solving complex problems in domains such as quantum chemistry, optimization, and machine learning, leveraging Variational Quantum Algorithms (VQAs) such as Quantum Approximate Optimization Algorithms (QAOA)~\cite{farhi2014quantum,qaoa2}, Variational Quantum Eigensolvers (VQE)~\cite{vqe1,vqe2}, and recently, quantum neural networks (QNNs)~\cite{schuld2019quantum,killoran2019continuous} as a hybrid quantum-classical framework in the Noisy Intermediate-Scale Quantum (NISQ) era.

These algorithms leverage Parameterized Quantum Circuits (PQCs) and optimize their parameters iteratively to improve performance. The expressibility of a PQC depends on factors such as the number of qubits and the circuit's depth. While greater expressibility increases the potential to solve complex problems, it comes with a No-Free-Lunch theorem: The more general and expressive the model, the reduced likelihood of successfully training the model. The trainability of quantum circuits diminishes due to challenges such as the presence of exponentially many local minima~\cite{you2021exponentially} and the Barren Plateau phenomenon~\cite{mcclean2018barren}. Specifically, the Barren Plateau refers to the exponential vanishing of gradients as the circuit depth or the number of qubits grows, making parameter optimization extremely difficult. Moreover, local minima in the optimization landscape can trap optimization trajectories, further complicating the scalability and effectiveness of VQAs.

Therefore, balancing trainability and expressibility is a key challenge in quantum circuit design. While circuits with high expressibility can explore complex solution spaces, they often require increased depth or parameterization, exacerbating the risk of barren plateaus and noise-related degradation in NISQ devices. 
Recent research has explored various strategies to address these issues. 
For example, it has been shown that using problem-inspired or hardware-efficient ansatze can mitigate the likelihood of encountering barren plateaus by constraining the optimization landscape to more trainable regions~\cite{grant2019initialization, sim2019expressibility}. Another approach involves optimizing the initialization of parameters to ensure gradients are sufficiently large during the early stages of training. Techniques such as layerwise training~\cite{skolik2021layerwise} and parameter initialization schemes based on symmetry considerations~\cite{pesah2021absence} have been proposed to achieve this.

Local cost functions, selective parameter training, and structured initialization methods have shown promise in mitigating trainability challenges without significantly compromising circuit expressibility. Moreover, techniques like symmetric pruning~\cite{symPrune}, which leverage circuit symmetries to reduce the effective parameter space, have demonstrated faster convergence and improved optimization performance. These advancements highlight the importance of reducing circuit complexity while preserving the circuit's ability to represent the desired quantum states.

\begin{figure*}[ht!]
    \centering
     \subfigure[Fully random activation]{\includegraphics[width=0.55\columnwidth]{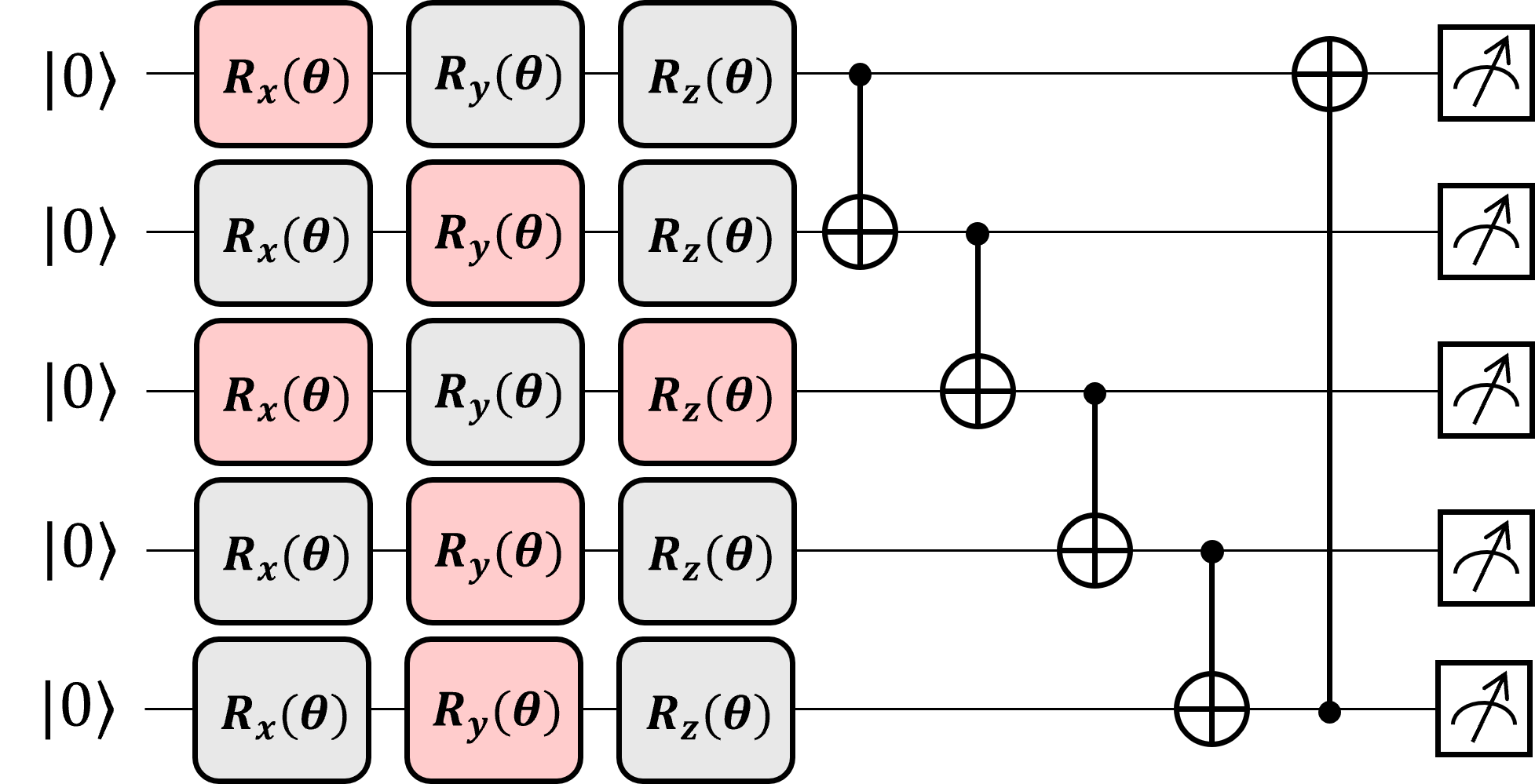}}
     \subfigure[Gate random activation]{\includegraphics[width=0.55\columnwidth]{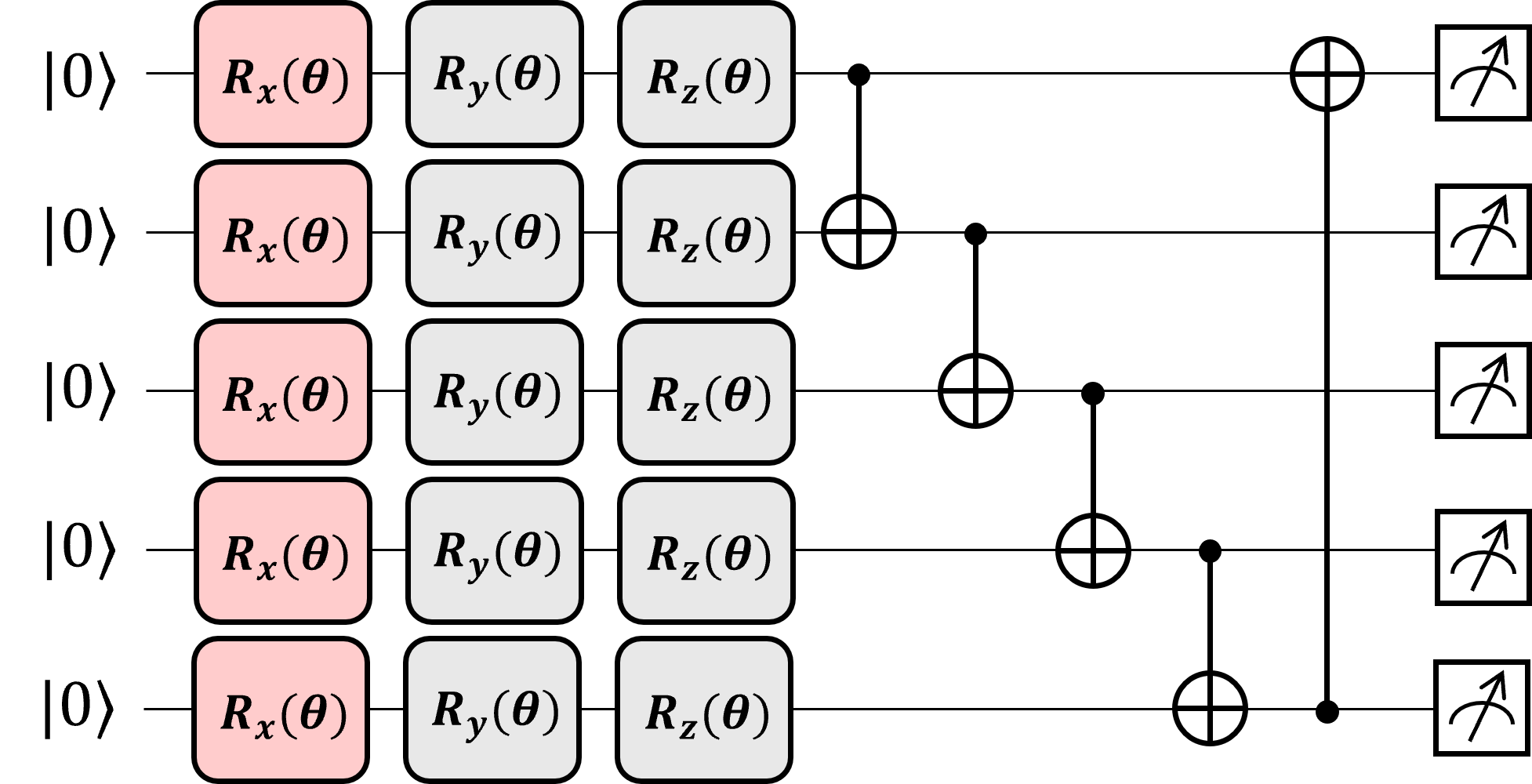}}
     \subfigure[Magnitude-based activation]{\includegraphics[width=0.9\columnwidth]{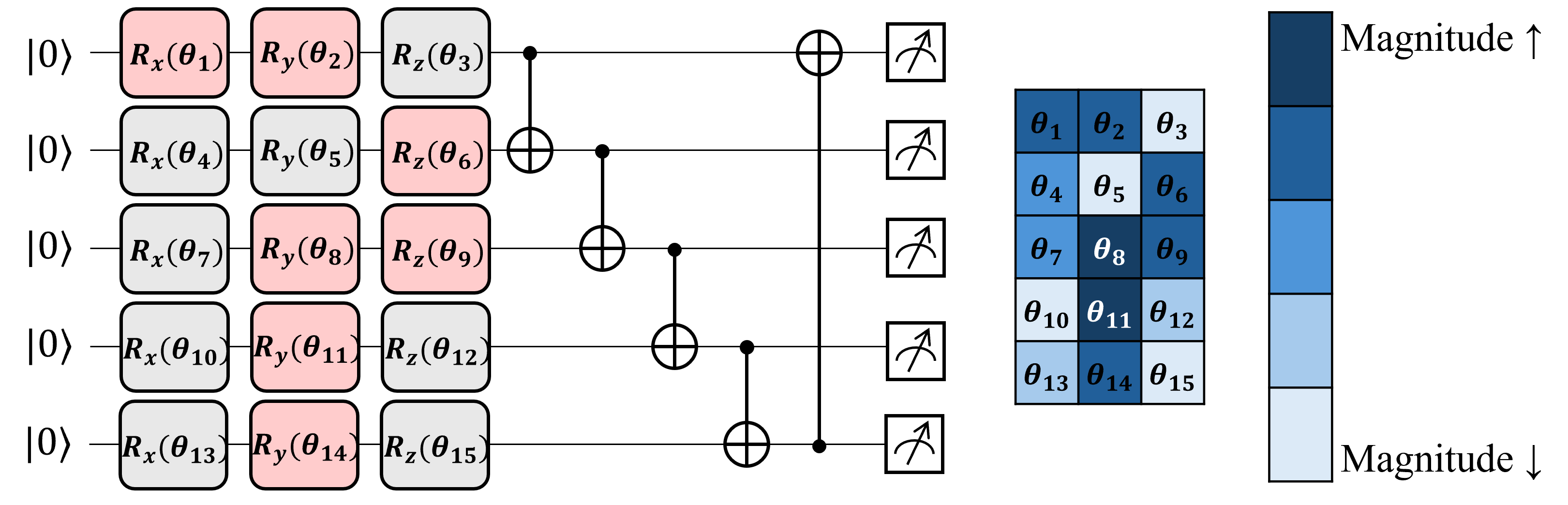}}
    \caption{An illustration of (a) fully random activation, (b) gate random activation strategy, and (c) magnitude-based activation where red-colored gates indicate selected gate. In (b), the RX gate is selected as the activated gate type, and in (c), the parameters with darker blue have a larger magnitude.}
    \label{fig:act_strat}
\end{figure*}

In this work, we focus on selective gate activation strategies as a practical and scalable solution to trainability challenges in quantum circuits. Specifically, we investigate the magnitude-based gate activation strategy, a novel approach that prioritizes gates with larger parameter magnitudes for activation. Unlike traditional methods that activate all gates simultaneously, selective gate activation introduces additional flexibility, allowing circuits to maintain expressibility while improving trainability. Additionally, this approach demonstrates resilience to noise, making it particularly suitable for real-world implementations on NISQ hardware.

This paper provides a detailed exploration of the magnitude-based gate activation strategy and its impact on circuit trainability and performance. We compare this approach with other activation strategies, such as fully random activation and gate-type random activation, and validate its effectiveness through experimental results. Our analysis considers various metrics, including the gap between exact and obtained ground state energies, fluctuations during training, and robustness across different gate activation percentages and warm-up iterations. 

For the remaining parts of the paper, we first provide background information about VQAs and the circuit trainability and expressibility and explain circuit gate activation strategies. We analyze each strategy with VQE experiments on 10 qubits Hamiltonian.


%% file: main.tex
In this section, we introduce various gate activation strategies to explore the relationship between circuit expressibility and the number of activated gates. We first provide background research related to circuit expressibility, followed by detailed information about gate activation strategy.

\subsection{Background}
\paragraph{Variational Quantum Algorithm}
VQAs are hybrid quantum-classical frameworks that gained popularity for solving problems in various domains such as quantum chemistry, optimization, and machine learning. VQAs leverage PQCs to approximate solutions to target problems by iteratively optimizing a cost function using classical methods~\cite{cerezo2021variational}. Among the VQAs, the VQE is the most widely studied area, which is tailored for finding the ground-state energy of quantum systems. When $U(\theta)$ is a parameterized unitary, that generates the quantum state $|\psi(\theta)\rangle = U(\theta)|\psi_0\rangle$ with initial quantum state $|\psi_0\rangle$, the parameters are optimized via minimizing the following loss function.
\begin{equation}
    l(\theta) = \langle \psi(\theta)|H|\psi(\theta)\rangle
\end{equation}
where $H$ is target Hamiltonian to solve.
VQE has demonstrated its potential in applications such as molecular electronic structure calculations and quantum many-body physics, where it provides efficient approximations of ground states by optimizing a Hamiltonian-dependent cost function \cite{mcclean2016theory}.

\begin{figure*}[!ht]
    \centering
     \subfigure[Fully random activation]{\includegraphics[width=0.33\textwidth]{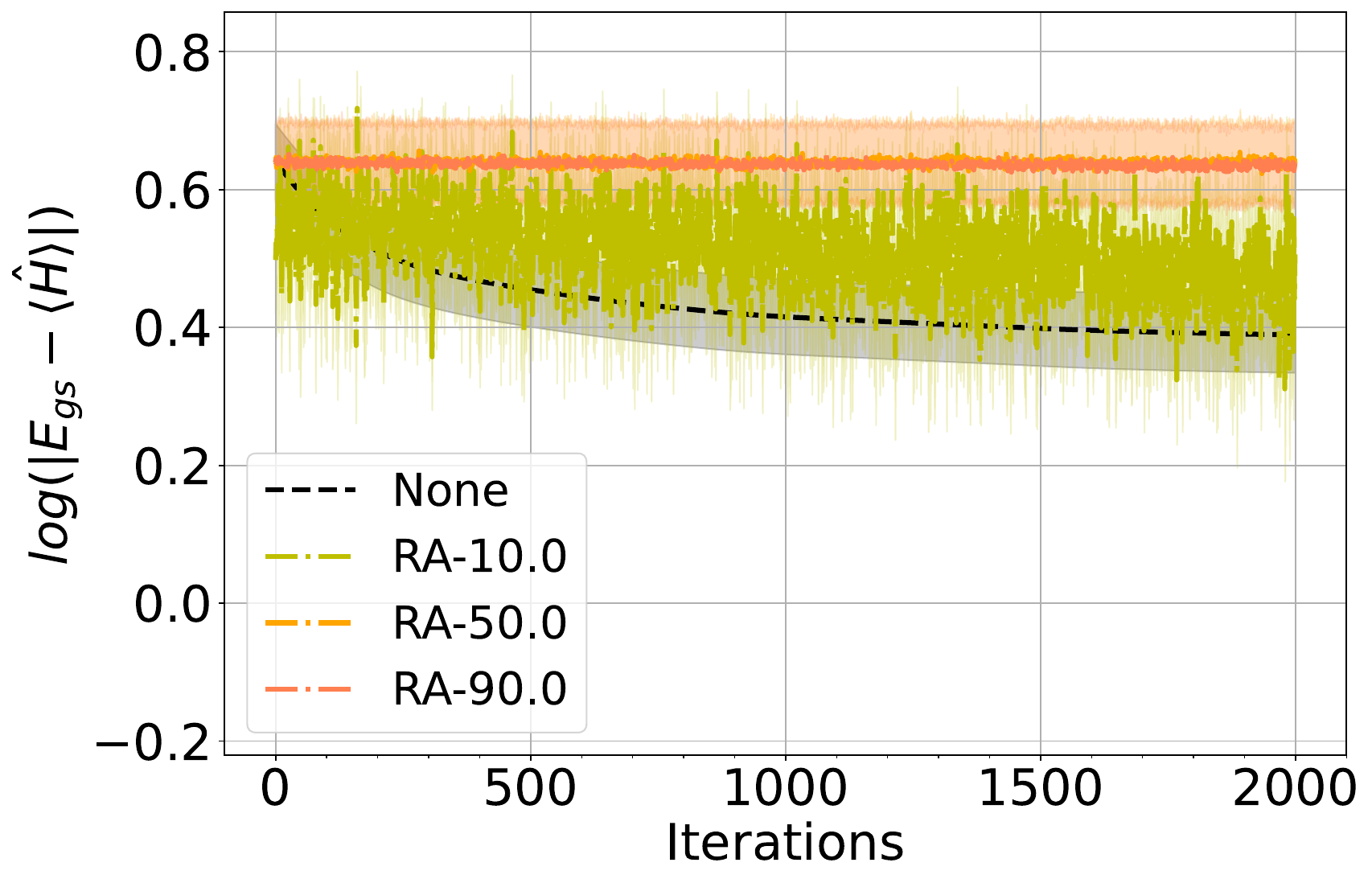}}
     \subfigure[Gate random activation]{\includegraphics[width=0.33\textwidth]{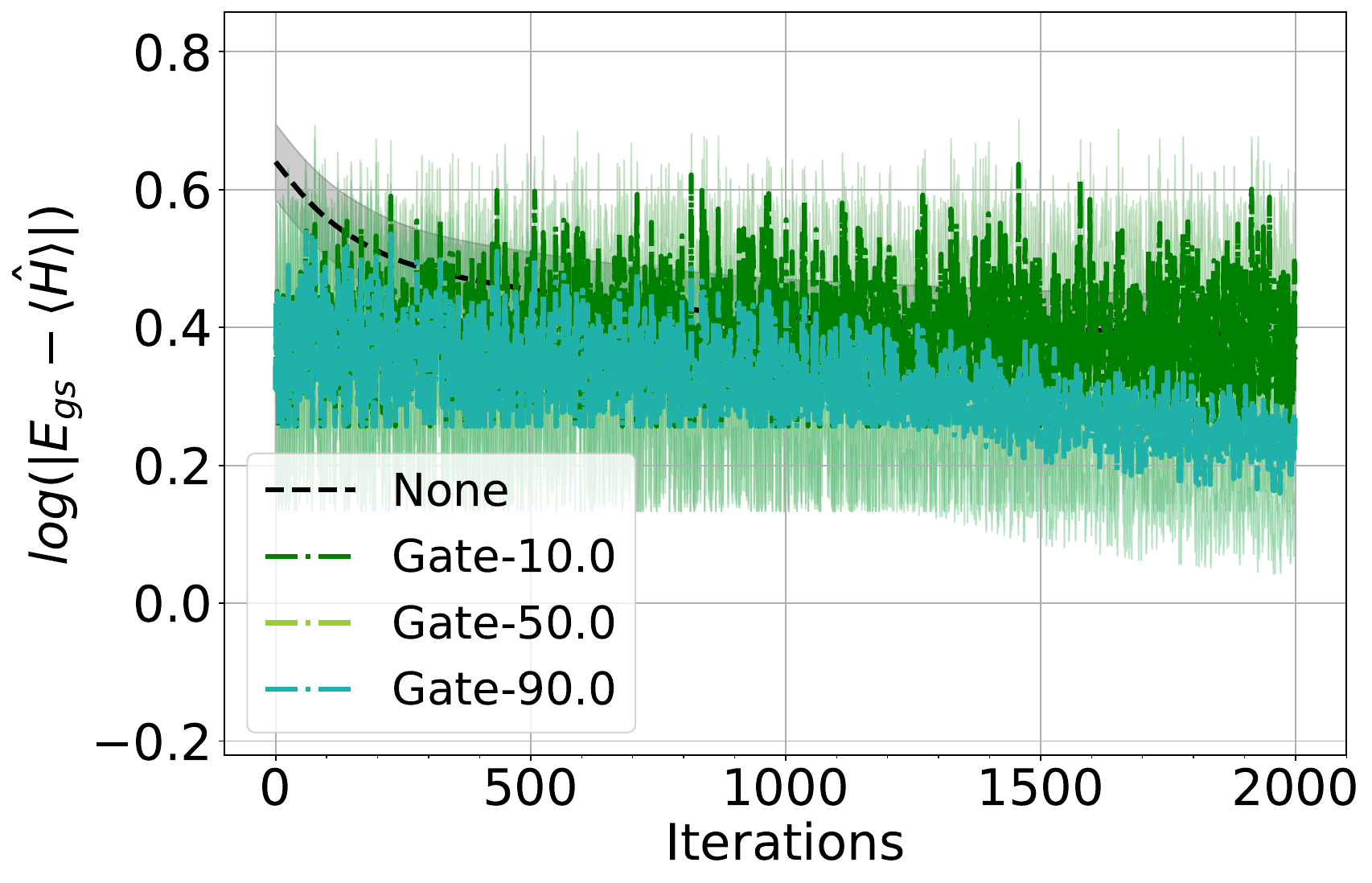}}
     \subfigure[Magnitude-based activation]{\includegraphics[width=0.33\textwidth]{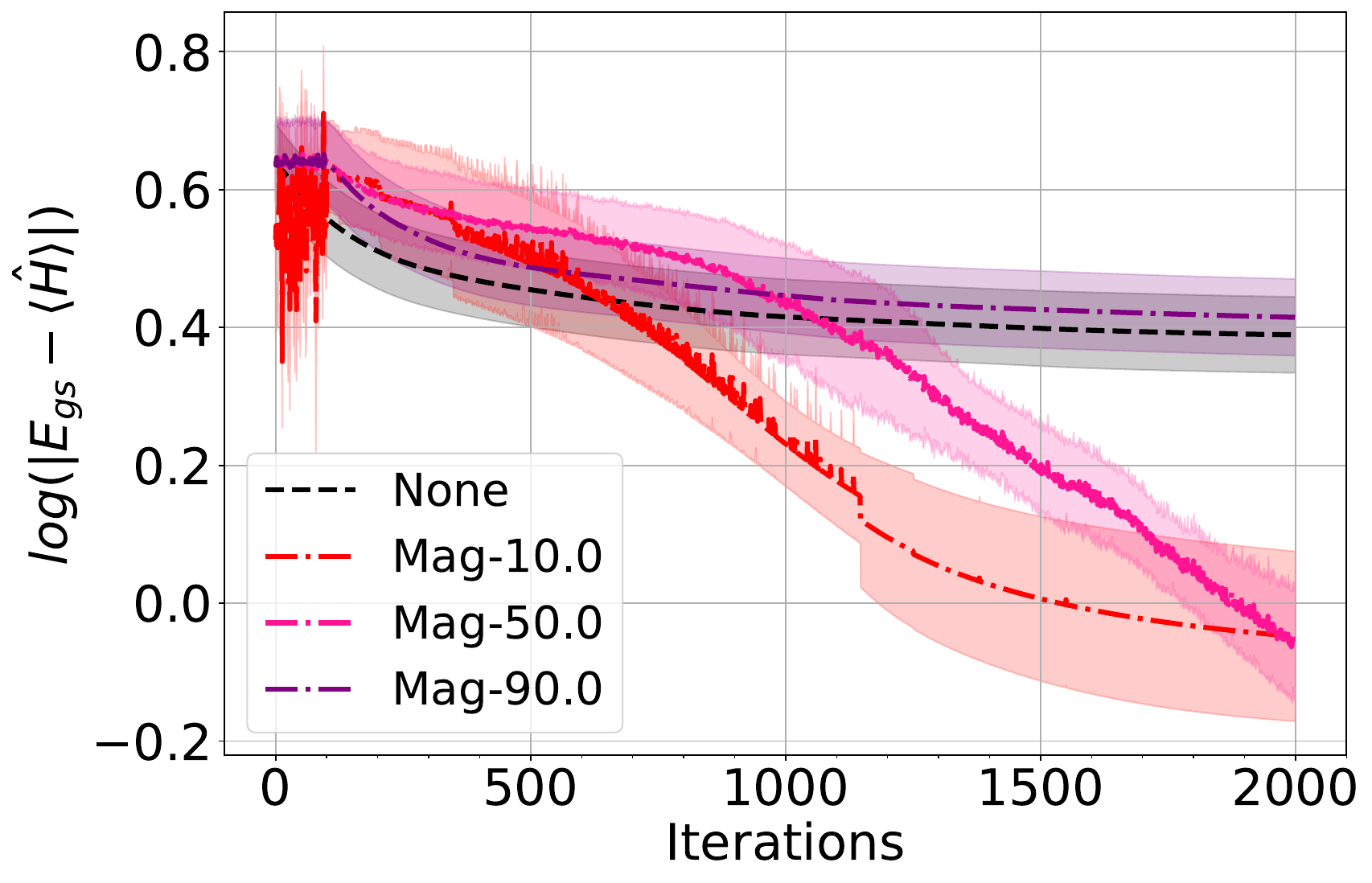}}
    \caption{Performance of strategies when $k$ varies from 10, 50, and 90. The shaded regions indicate one standard deviation ($[\mu-\sigma/2, \mu+\sigma/2]$). \textit{RA} failed to train as more gates were activated, whereas \textit{Mag} remained trainable even with up to $50\%$ gate usage.}
    \label{fig:mod_k}
\end{figure*}
\paragraph{Circuit Trainability and Expressibility}
Circuit trainability and expressibility are two fundamental aspects of quantum circuit design that must be carefully balanced to ensure the effective application of VQAs. \textbf{Trainability} refers to the ease with which circuit parameters can be optimized to minimize a given cost function, typically using gradient-based methods. Poor trainability often arises due to phenomena like the Barren Plateau problem, where gradients vanish exponentially with the number of qubits or circuit depth, severely hindering the optimization process~\cite{mcclean2018barren}.
The trainability of the quantum circuit can be measured via various metrics, one common way is to measure gradient magnitudes which is well-described in~\citet{holmes2022connecting}. The gradient of the loss function corresponding to the parameter $\theta_i$ is determined using partial derivative  $\partial_i l = \frac{\partial l(\theta)}{\partial\theta_i}$. Using Chebyshev inequality, the probability that the partial derivative of the cost deviates from its average of zero is as follows:
\begin{equation}
    P(|\partial_i l| \geq \delta) \leq \frac{Var[\partial_i l]}{\delta^2}
\end{equation}
where the variance is $Var[\partial_i l] = \langle(\partial_i l)^2 \rangle - \langle\partial_i l\rangle^2$.

On the other hand, \textbf{expressibility} measures the ability of a quantum circuit to represent a wide range of quantum states. Highly expressible circuits can capture complex solution spaces, but this often comes at the cost of deeper circuits or more parameters, exacerbating trainability issues by increasing the likelihood of Barren Plateaus~\cite{sim2019expressibility}.

A well-designed quantum circuit must balance these competing factors: while increased expressibility allows the circuit to solve more complex problems, excessive depth or parameterization may render the circuit untrainable. Recent studies have proposed strategies like local cost functions~\cite{cerezo2021variational}, selective parameter training~\cite{volkoff2021large}, and structured initializations~\cite{grant2019initialization} to mitigate trainability challenges without compromising expressibility. Additionally, the Random Activation (RA) strategy is proposed by~\citet{liu2023training}, where RA improves trainability by progressively activating subsets of gates, mitigating Barren Plateaus, and avoiding local minima through reduced parameter dimensionality and increased gradient visibility during early training.

\subsection{Gate Activation Strategy}
We categorize the gate activation strategies into three types: fully random activation, gate-type random activation, and magnitude-based activation. Only the gates that are activated are updated and others are not. An intuitive illustration of each activation strategy is depicted in Figure~\ref{fig:act_strat}. Instead of increasing the number of activated gates progressively, we activated a fixed number of gates per iteration.
\paragraph{Fully random activation} Gates are activated without regard to their type or the magnitude of their parameters. In each iteration, $k\%$ of the gates are randomly selected for activation.
\paragraph{Gate random activation} Gates are randomly selected from a single type of rotation gate. First, the gate type will be chosen among RX, RY, and RZ gates (representing rotations around the X, Y, and Z axes, respectively). Within the selected gate type, gates will be randomly activated. If the total number of selected rotation gates is less than $k\%$ of all gates, no additional gates from other types are activated.
\paragraph{Magnitude-based activation} Gates with larger magnitudes are prioritized for activation. In classical neural networks, neurons with higher activation outputs are typically regarded as more important. Inspired by this, gates with larger magnitudes are chosen for activation. Additionally, gates with smaller parameters induce minimal rotation of qubits, suggesting they have a lesser impact on the overall system. We select top $k\%$ parameters to activate the same as other strategies.

%% file: exp.tex
We carry out experiments to explore the performance of VQE with different activation strategies and validate the effectiveness of the Magnitude-based gate activation strategy. 

\subsection{Experiment Settings}

\begin{table}[hbt]
\centering
\resizebox{\columnwidth}{!}{
\begin{tabular}{>{\centering\arraybackslash}m{2.5cm} *{5}{>{\centering\arraybackslash}m{1.2cm}}}
\toprule
                                 & $C_2$ & $HF$ & $LiH$ & $Li_2$ & $OH^{-}$ \\ \midrule
Charge                           & 0           & 0         & 0         & 0           & -1           \\
Active Electrons                 & 8           & 8         & 2         & 2           & 8            \\
Bond Length (\AA)                & \multicolumn{5}{c}{0.5, 0.7, 0.9, 1.1, 1.22, 1.3, 1.5, 1.7, 1.9, 2.1, 2.3, 2.5} \\ 
Active Orbitals                  & \multicolumn{5}{c}{5} \\ \bottomrule
\end{tabular}
}
\caption{Parameters for generating 10 qubits molecule Hamiltonian.}
\label{tab:mol_ham_params_10}
\end{table}

In this subsection, we provide information about experiments, such as a problem to solve, and experiment environments.

\paragraph{Target Hamiltonian} We select Molecule Hamiltonian as a target Hamiltonian to solve, where Hamiltonian is generated via the Pennylane function. The list of molecules and bond length to generate Hamiltonian is provided in Table~\ref{tab:mol_ham_params_10}. We randomly sampled 6 different Hamiltonian to verify the average performance of the strategy. The fermionic Hamiltonian can be expressed as a linear combination of tensor products of Pauli operators using the Jordan-Wigner transformation~\cite{seeley2012bravyi}, as shown below:
\begin{equation}
    H = \sum_{j}C_j\otimes_i \sigma_i^{(j)}
\end{equation}
where $C_j$ is a scalar coefficient and $\sigma_i$ represents an element of the Pauli group ${I,X,Y,Z}$.

\begin{figure}[ht]
    \centering
    \includegraphics[width=0.9\columnwidth]{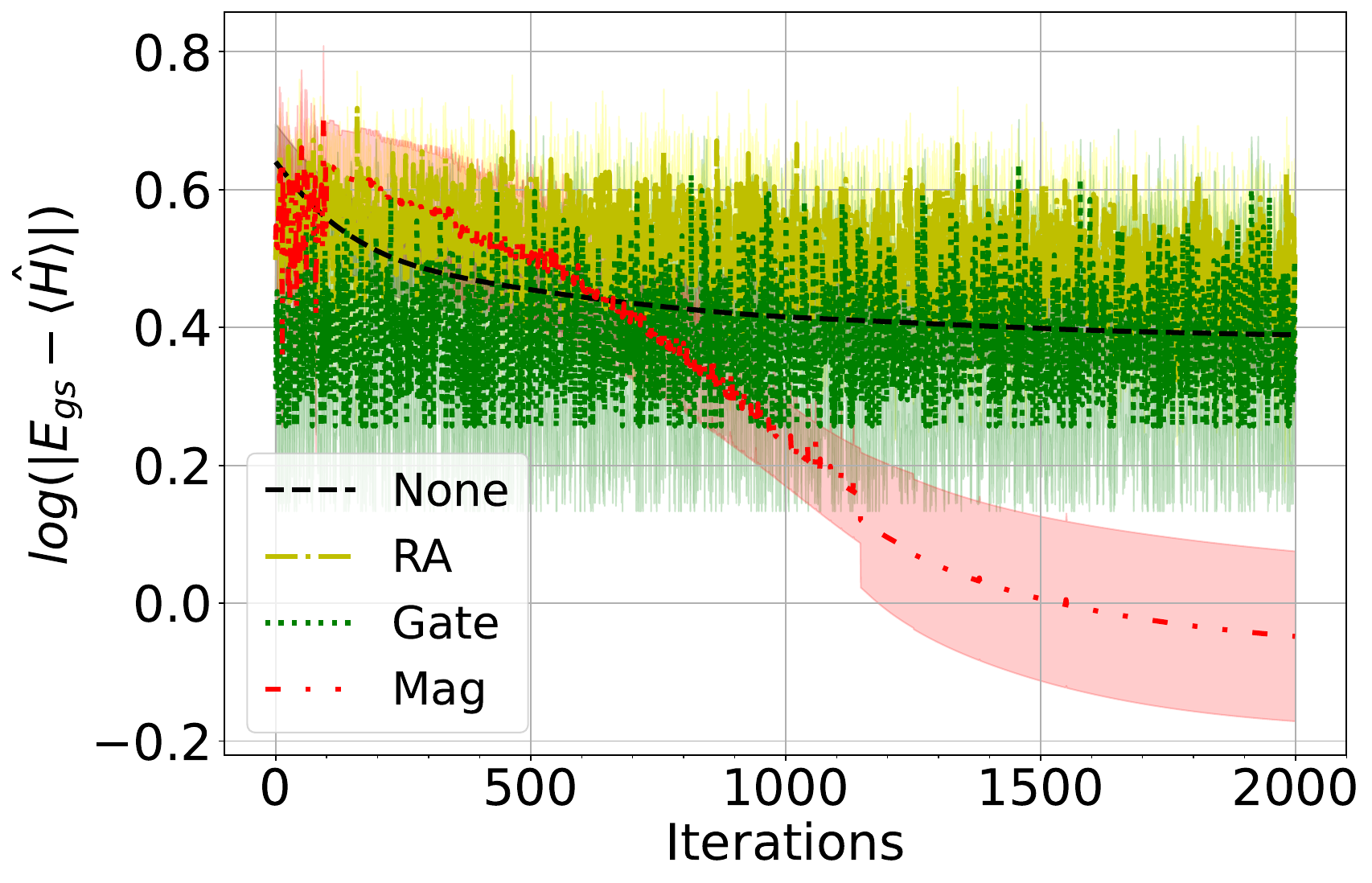}
    \caption{Performance of different strategies when $10\%$ of gates are selected. The shaded regions indicate one standard deviation ($[\mu-\sigma/2, \mu+\sigma/2]$).}
    \label{fig:overall_perf_0.1}
\end{figure}

\paragraph{Experiment Environment} The ansatz for training VQE is 7 layers of Strongly Entangling Layers~\cite{schuld2020circuit} which contains RX, RY, and RZ gate for every layer, and all qubits are entangled via CNOT gate (refer to Figure~\ref{fig:act_strat} for single layer design). We utilize 10 qubits to solve the target Hamiltonian, resulting in a total of 210 parameters for the circuit. The circuit is optimized over 2000 iterations using the Adam optimizer~\cite{kingma2014adam} with a learning rate of 0.001. It is worth mentioning that the experiment is conducted without noise and uses exact expectation values to evaluate the performance of gate activation clearly. 

We measure the gap between the exact ground state energy ($E_{gs}$) calculated via classical computer and calculated energy ($\langle\hat{H}\rangle$) obtained by VQE and take log-scale ($log(|E_{gs}-\langle\hat{H}\rangle|)$) to check the performance.

\subsection{Results}
We first compare three gate activation strategies: Random Activation (\textit{RA}), Gate-Random Activation (\textit{Gate-RA}), and Magnitude-Based Activation (\textit{Mag}), under the condition where only $10\%$ of the gates — equivalent to 21 gates — are selected. The warm-up iteration for \textit{Mag} is set to 100. 

In Figure~\ref{fig:overall_perf_0.1}, we can clearly tell that \textit{Mag} shows superior performance compared to others, with less fluctuation and with the smallest gap. An interesting observation is that \textit{Gate-RA} outperforms \textit{RA}, suggesting that training the same gates simultaneously enhances performance compared to mixing different gates.

We then compare the performance of each strategy when $k$ is modified from 10, 50, and 90 in Figure~\ref{fig:mod_k}. \textit{RA} struggles to train effectively as more gates are utilized, achieving performance comparable to standard VQE training only when 
$10\%$ of the gates are selected. In contrast, \textit{Gate} demonstrates some trainability depending on the number of gates; however, it exhibits significant fluctuations during training, indicating a poor ability to converge. On the other hand, \textit{Mag} consistently outperforms the other strategies across various gate selections and remains robust and adaptable, even when the number of gates increases.

\begin{figure}[ht]
    \centering
    \includegraphics[width=0.9\columnwidth]{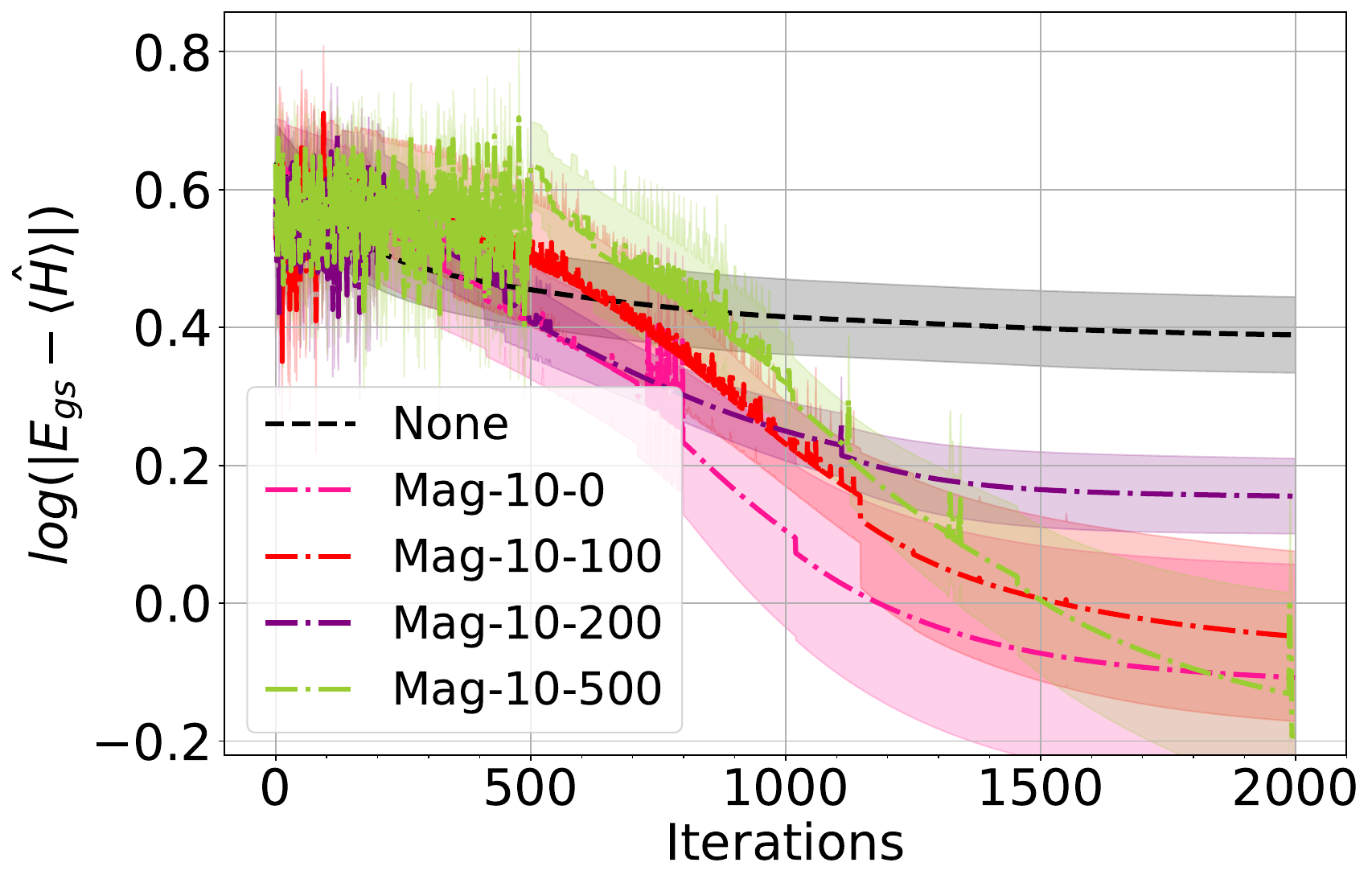}
    \caption{Convergence of \textit{Mag} with different warm-up iteration numbers, 0, 100, 200, and 500. The shaded regions indicate one standard deviation ($[\mu-\sigma/2, \mu+\sigma/2]$).}
    \label{fig:warmup_perf}
\end{figure}

Lastly, to evaluate the performance of \textit{Mag}, we vary the number of warm-up iterations, testing values of 0, 100, 200, and 500, while selecting the top $10\%$ of gates. Interestingly, the results reveal that the best performance is achieved when no warm-up iterations are used. This suggests that even with random initialization, gates with larger magnitudes have a significant impact on obtaining the expected values. Their influence appears sufficient to guide the training process onto the correct path without requiring additional warm-up stages.

%% file: conclusion.tex
\subsection{Conclusion}

We introduce the magnitude-based gate activation strategy that can enhance the trainability of the quantum circuit in VQE and provide experiments that show performance concerning the number of gates and warm-up iterations. 

Overall, the results suggest that selectively training gates can significantly enhance the trainability of quantum circuits by reducing the parameter search space. However, choosing which gates to train requires careful consideration, as inappropriate selection might limit the circuit's expressivity or hinder its ability to learn the desired function, as we can see from the failure of \textit{RA}. An optimal strategy must strike a balance between reduced parameterization and sufficient circuit complexity to ensure effective learning.

We believe that selective gate activation in complex circuits offers a promising approach to mitigating the Barren Plateau problem. By enhancing both expressivity and trainability, this method enables quantum circuits to address more challenging problems effectively.

\subsection{Future Direction}


In this work, we propose a magnitude-based gate activation strategy and assess its effectiveness through an evaluation of various gate activation approaches. This method addresses the challenges posed by the Barren Plateau problem, which is often tackled by identifying optimal initialization points. By introducing a fresh perspective, we aim to inspire new directions for research in this area. 

Looking ahead, this foundational strategy opens the door for significant advancements. For example, future research could focus on developing adaptive gate selection methods that dynamically adjust based on the training progress or the characteristics of the optimization landscape. Similarly, strategies that progressively activate additional gates, refining the circuit's expressiveness until convergence, present an exciting avenue for exploration.

Moreover, our findings reveal that gate-based random activation outperforms simple random activation. This insight suggests opportunities for further investigation into multi-qubit gate activation schemes or selective activation targeting non-parameterized gates. Such explorations could lead to a deeper understanding of circuit trainability and the development of more robust methodologies to overcome the limitations imposed by Barren Plateaus.